\documentclass[aps,prl,twocolumn,letterpaper,10pt,footinbib,superscriptaddress,floatfix,amsfonts,amsmath,amssymb]{revtex4-2}

\usepackage[english]{babel} 
\usepackage[utf8]{inputenc}
\usepackage[T1]{fontenc}
\usepackage{amsmath}
\usepackage{amssymb}
\usepackage{amsthm}
\usepackage{mathtools}
\usepackage{geometry}
\usepackage{bbold}
\usepackage{graphicx}
\usepackage{caption}
\usepackage{subcaption}
\usepackage{physics}
\usepackage{listings}
\usepackage{float}
\usepackage{floatflt}
\usepackage{blindtext}

\usepackage{bm,physics,graphicx}
\usepackage[utf8]{inputenc}
\usepackage[english]{babel}
\usepackage[dvipsnames]{xcolor}
\graphicspath{{./Figures/}}

\begin{document}
\title{Vortex Excitations of Dirac Bose-Einstein Condensates}
\author{Joris Schaltegger}
\affiliation{Nordita,
KTH Royal Institute of Technology and Stockholm University
106 91 Stockholm, Sweden}
\author{A.V. Balatsky}
\affiliation{Nordita,
KTH Royal Institute of Technology and Stockholm University
106 91 Stockholm, Sweden}
\affiliation{Department of Physics, University of Connecticut, Storrs, Connecticut 06269, USA}
\date{\today}
\maketitle

\section*{Abstract}
We explore vortices in non-equilibrium Dirac Bose-Einstein condensates (Dirac BEC) described by a stationary Dirac Gross-Pitaevskii equations (GPE). We find that the multi-component structure of Dirac equation enables the difference in phase winding of two condensates with respective phase winding number differing by one, $\ell_a - \ell_b = \pm 1$. We observe three classes of vortex states distinguished by their far-field behavior: A ring soliton on either of the two components in combination with a vortex on the other component, and, in the case of strong inter-component interactions, a vortex profile on both components. The latter are multiple core vortices due to the phase winding difference between the components. We also address the role of a Haldane gap on these vortices, which has a similar effect than inter-component  by making the occupation on either sublattice more costly. We employ a numerical shooting method to reliably identify vortex solutions and use it to scan large parts of the phase space. We then use a classification algorithm on the integrated wavefunctions to establish a phase diagram of the different topological sectors.

\section{Introduction} \label{sec:intro}
Dirac nodes in the excitation spectrum of a condensed matter system are the consequence of additional symmetries present, for example sublattice symmetry in graphene and time-reversal symmetry in topological insulators. While most of the time, fermionic Dirac and Weyl materials hosting electrons are studied, the notion of Dirac materials can easily be extended to systems with bosonic quasi-particles obeying analogous symmetries \cite{Wehling2014}. Indeed, Dirac or Weyl points have been observed in a wide variety of bosonic excitations, for example in photonic crystals \cite{Lu2015,Wang2016,Yang2019}, exciton-polariton condensates \cite{kim_2013,jacqmin_2014}, and hexagonal magnets \cite{Samuelson71,Yelon71,Fransson2016,pershoguba_2018,Chen2018,Xu2018,mcclarty_2018}. The defining feature of Dirac materials is the existence of low-energy excitations that are described by the Dirac equation rather than the Schrödinger equation. In fermionic systems, it is possible to tune the Fermi level to the Dirac nodes, for example via doping, which means that Dirac-like excitations can be studied in or near thermal equilibrium. In contrast, the equilibrium chemical potential for bosonic systems is close to zero at low temperatures, i.e. the ground state is characterized by the absence of bosonic excitations. Maintaining a non-zero chemical potential for bosonic degrees of freedom requires a continuous external drive, just as is the case for other non-equilibrium condensates where bosons are produced by an external source, typically optical or microwave light. We will refer to these cases of condensates as pumped or non-equilibrium condensates. The pumping of bosonic quasi-particles and subsequent condensation is well-established for magnon \cite{demokrit_2006,demidov2007,Demidov2008,Serga2014,Nowik-Boltyk2012,Bozhko2016,Bozhko2019,Borisenko2020} and exciton-polariton condensates \cite{Eisenstein2004,Kasprzak2006,balili2007,deng2010}. Such non-equilibrium condensates are typically generated at the minimum of the dispersion, where the lifetime of bosonic quasiparticles is expected to be the longest. Recent experiments on thin-film yttrium iron garnets additionally observe a magnon condensate at higher energy bands \cite{bailey2022}. In light of this development, the notion of a pumped condensate of Dirac bosons, first introduced by Sukhachov et.al. \cite{sukhachov2020boseeinstein}, seems plausible.



In this paper, we explore the nature of vortices in Dirac BEC. As a specific example, we focus on the case of two-dimensional, hexagonal magnets, such as $\text{Cr}\text{I}_3$, where Dirac nodes appear in the magnon spectrum \cite{}, but the results about multiple core components and structure of Dirac BEC vortices is general and applicable to other systems that support Dirac BEC. Following Ref. \cite{sukhachov2020boseeinstein}, we use a phenomenological model for the free energy density of a Dirac BEC, taking into consideration the spinor structure and linear kinetic energy of the underlying Dirac equation. In our minimal model, we keep only the most important interaction terms and restrict to stationary solutions of the resulting Dirac Gross-Pitaevskii equations. We observe the following:
\begin{enumerate}
    \item The linear kinetic energy of the Dirac equation leads to the unusual case of {\em different winding numbers} for the two sublattice components, related by $\ell_a = \ell_b \pm 1$.
    \item As the consequence of  different inter- vs intra-component interactions we observe three distinct classes of topological excitations characterized by their far-field behavior: The first two types consist of configurations with a locally confined ring soliton on one sublattice, in combination with a vortex configuration on the other sublattice. Which of the two components is dominant depends on the relative interaction strengths and the Haldane gap. The third type is characterized by a vortex configuration for both components, which appears for sufficiently strong inter-component interactions. This class of solutions is of particular interest because the angular momentum jump results in multiple cores for the two components.
    \item To reach these conclusions we employ a numerical shooting algorithm to scan large parts of the phase space spanned by the three independent model parameters. We then use an cluster algorithm on the integrated radial wavefunctions to compute the phase diagram of topological excitations.
    \item We find the phase diagram for multi-component Dirac BEC that shows that the condensate with the lower winding number is favoured, which is consistent with the lower rotational energy. However, tuning the Haldane gap and lowering the interaction strength on the recessive sublattice can invert the occupation of the condensate components. For sufficiently large inter-component interactions, a multiple core vortex can occur. The Haldane gap and decreasing the repulsion on the recessive component facilitates the onset of the mixed phase.
    \item The phase diagrams are qualitatively similar for the topological sectors $\ell_a = 0,1 (\ell_b = 1,2)$.

\end{enumerate}

The structure of this paper is as follows: After the introduction in section 1 we discuss a minimal model and the general setup for the analysis of vortices in Dirac BEC in section 2. In section 3 we present the results for the topologically distinct vortex states present in the Dirac BEC. In section 4 we give an outlook and conclusion. We describe the numerical methods and shooting algorithm for solving the Dirac GPE in Appendix A.

\section{Minimal model of Dirac BEC} 
\label{sec:model}

Following Ref. \cite{sukhachov2020boseeinstein}, we use a phenomenological ansatz for the free energy of the Dirac BEC,
\begin{widetext}
\begin{align}
	\label{free_energy}
		\mathcal{F} &= (c_0 -\mu) n_{tot} + \Delta \sum_{\zeta = \pm} \big(n_{a,\zeta} - n_{b,\zeta}\big) - iv\Psi^\dagger \big(\tau_z \otimes \vec{\sigma} \cdot \vec{\nabla} \big) \Psi \nonumber \\
		&+ \sum_{\zeta = \pm} \big[\frac12 g_a n_{a,\zeta} n_{a,\zeta} + \frac12 g_b n_{b,\zeta}n_{b,\zeta} + g_{ab} n_{a,\zeta} n_{b,\zeta}
		\big].
\end{align}
\end{widetext}

Here, the wavefunction is expressed as the spinor $\Psi = \begin{pmatrix}
	\psi_{a,+} & \psi_{b,+} & \psi_{a,-} & \psi_{b,-}
\end{pmatrix}^T$, where $a,b$ is the sublattice (or, more generally, pseudo-spin) index, and $\zeta=\pm$ is the valley index. The condensate densities are denoted by $n_{j,\zeta} = \psi_{j,\zeta}^\dagger \psi_{j,\zeta}$, and $n_{tot} = \Psi^\dagger \Psi$ is the total density. Furthermore, $\vec{\sigma}$ is the vector of Pauli matrices, $\tau_z$ is the Pauli matrix in the valley degree of freedom, $c_0$ is the energy of the Dirac nodes, $\mu$ is the effective chemical potential generated by the external drive. To keep the discussion general, we introduce the Haldane gap $\Delta$. The velocity $v = \left|\frac{d\epsilon}{d\vec{k}}\right|_{\vec{k}_D}$ is the equivalent to the Fermi velocity at the Dirac nodes. Here, the kinetic terms are linear in derivatives due to the expansion around the Dirac point of the energy spectrum. Note that we have restricted to interactions between sublattices; the valley degrees of freedom are completely decoupled in our model. We remark that interactions between different valleys can for example occur from microscopic Kitaev interactions. For simplicity the valley index will be dropped from now on. The origin of inter-component pseudo-spin interactions can be motivated by calculating the next-to-leading order Holstein-Primakoff transform of a Heisenberg ferromagnet on a honeycomb lattice \cite{pershoguba_2018}. An intra-component repulsion is required for the spatial stability of the condensate \cite{Kourakis2005}. In magnon condensates in YIG, the observed repulsion is likely a consequence of additional dipole interactions for non-uniform densities \cite{Borisenko2020}.

The Dirac Gross-Pitaevskii equations follow from the free energy density by variation of the spinor components,
\begin{widetext}
\begin{subequations}
	\label{gross_pitaevskii_eq}
	\begin{align}
		i\partial_t \psi_a &= \left(c_0 -\mu + \Delta \right)\psi_a - iv\left( \partial_x -i\partial_y \right) \psi_b + g_a n_a \psi_a + g_{ab} n_b \psi_a, \\
		i\partial_t \psi_b &= \left(c_0 -\mu - \Delta \right)\psi_b - iv\left( \partial_x +i\partial_y \right) \psi_a + g_b n_b \psi_b + g_{ab} n_a \psi_b.
	\end{align}
\end{subequations}
\end{widetext}

In the presence of an external drive, the Dirac GPE should be extended to take into consideration the source and decay terms for the Dirac bosons. However, here we assume a stationary state where the source and decay terms cancel each other exactly. Homogeneous solutions of the stationary GPE have been discussed in Ref. \cite{sukhachov2020boseeinstein}, where it is shown that, depending on the model parameters, the ground state either has a non-zero density for both pseudo-spins, or one of the components vanishes.

We now discuss spatially non-uniform vortex solutions of the Dirac GPE, which are an example of a self-interacting non-linear Dirac equation. This class of differential equations has been studied extensively in the past, see e.g. Ref. \cite{haddad1,haddad2,haddad3,haddad4,haddad5,haddad6,haddad7,poddubny2018,cuevas2016,BOUSSAID2016}. While the focus has mostly been on locally bound solitons, we instead aim to find vortex solutions that recover to one of the homogenous ground states outlined above in the far-field limit. We employ the vortex ansatz
\begin{equation}
	\label{vortex_ansatz}
	\psi_{a(b)}(r,\theta) = \rho_{a(b)}(r) e^{i \ell_{a(b)} \theta}.
\end{equation}
The winding numbers $\ell_{a(b)}$ must be integers to ensure single-valued wavefunctions, and dictate the behavior at the vortex core, $\rho_{a(b)}(r) \propto r^{\ell_{a(b)}} \ (r \rightarrow 0)$.

It is convenient to rewrite the Dirac GPE in polar coordinates, which decouples the angular and radial parts. The linear kinetic terms yield a factor of $e^{\pm i \theta}$, from which the condition $\ell_a = \ell_b - 1$ for the winding numbers follows immediately. Decoupling the angular part gives the radial equations
\begin{widetext}
\begin{subequations}
	\label{gpe_radial}
	\begin{align}
		\Big[g_a \big(n_a - n_{a,\infty}\big) + g_{ab} n_b \Big] \rho_a &= iv \left( \partial_r + \frac{l_b}{r} \right) \rho_b \\
		\Big[g_b \big(n_b - n_{b,\infty}\big) + g_{ab} n_a \Big] \rho_b &= i v\left( \partial_r + \frac{1-l_b}{r} \right) \rho_a,
	\end{align}
\end{subequations}
\end{widetext}

where we have introduced the effective parameters $n_{a(b),\infty} \vcentcolon= \frac{\mu - c_0 \mp \Delta}{g_{a(b)}}$ that can be though of as far-field densities of condensate components. The usefulness of these parameters becomes obvious when calculating the far-field limits of the condensates. Indeed, $|\rho_{a(b)}|^2 \xrightarrow{r \rightarrow \infty} n_{a(b),\infty} \ \text{or} \ 0$ in the absence of inter-component interactions ($g_{ab} = 0$). We remark that changing the winding number $\ell_b\rightarrow 1-l_b$ effectively switches the sublattices, hence it is sufficient to consider $\ell_b\geq1$. Furthermore, the case $\ell_b = 1, \ell_a = 0$ is special because only one pseudo-spin has a non-zero phase winding. Our numerical methods are less reliable in this topological sector, hence we focus on the $\ell_b = 2, \ell_a = 1$ sector.

Identifying vortex solutions of \eqref{gpe_radial} is achieved by a numerical shooting algorithm \cite{haddad6}: From a Taylor expansion around the vortex core, it follows that the leading Taylor coefficient of $\rho_a$ is a free parameter. A vortex solution is obtained by tuning this initial condition such that oscillations in the far field are suppressed. Details of the shooting algorithm are outlined in Appendix A.

We automated this process by developing a bisection algorithm that minimizes the number of oscillations of the integrated wavefunction. This allows us to scan large parts of the phase space spanned by the model parameters. In practice, we rescale the radial coordinate to $r\rightarrow \frac{\mu - c_0 - \Delta}{v} r$ and the density to $\rho_{a,b} \rightarrow \frac{1}{\sqrt{n_{a,\infty}}} \rho_{a,b}$. This reduces the number of independent model parameters to three, namely the relative interaction strengths $\frac{g_b}{g_a}$ and $\frac{g_{ab}}{g_a}$, and the effective Haldane gap $\tilde{\Delta} = \frac{c_0 - \mu - \Delta}{c_0 - \mu + \Delta}$.

\section{Results} \label{sec:results}

\subsection{$\ell_b = 2, \ell_a = 1$ sector}

\begin{figure*}
	\centering
	\includegraphics[width=0.9\textwidth]{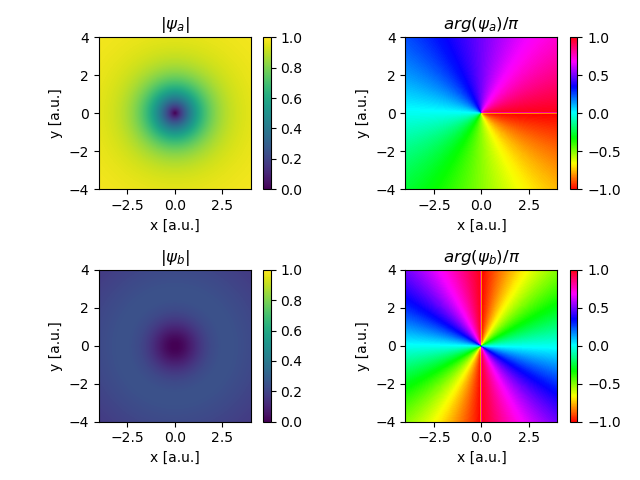}
	\caption{Normalized condensate density (left panels) and phase (right panels) of the A-vortex for $\ell_a=1,\ell_b=2,\frac{g_b}{g_a} = 1,\frac{g_{ab}}{g_a} = 0, \tilde{\Delta} = 1$. Upper (lower) panels show density and phase for sublattice A(B).}
	\label{fig:l2_vortex_a}
\end{figure*}

In the absence of inter-component interactions, $g_{ab} = 0$, we observe vortices with finite far-field density for one component, and a ring-soliton profile on the other component. In the totally symmetric case, population of sublattice A is preferred. This is consistent with the lower kinetic energy, which is proportional to the square of the phase winding number, $E_{kin} \sim \ell_{a(b)}^2 \ln \frac{L}{\xi}$ for $\rho_{b(a)} \rightarrow 0$. Here, $\xi$ is the healing length of the vortex and $L$ is the system size. The logarithmic divergence is an artifact from the presence of an unpaired vortex; introducing an anti-vortex would lift the divergence. The density and phase of such a configuration is shown in figure \ref{fig:l2_vortex_a}. The density plots show the vortex profile on component A (upper left) and a ring-soliton on component B (lower left), while the phase corresponds to winding numbers $\ell_a = 1$ (upper right) and $\ell_b = 2$ (lower right). We call this configuration an A-vortex excitation.

\begin{figure*}
	\centering
	\includegraphics[width=0.9\textwidth]{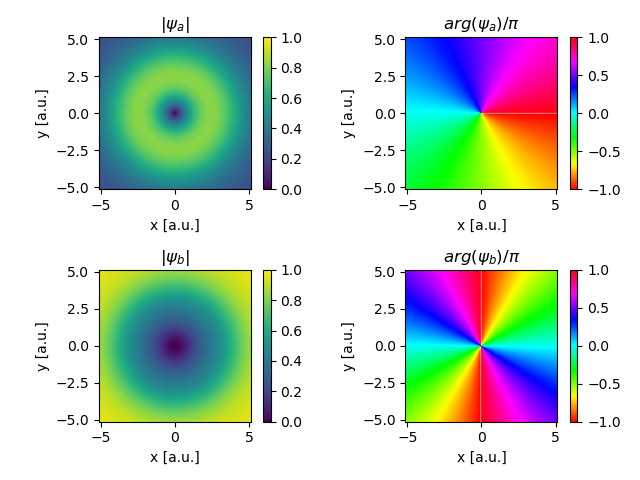}
	\caption{Normalized condensate density (left panels) and phase (right panels) of the B-vortex for $\ell_a=1,\ell_b=2,\frac{g_b}{g_a} = 1,\frac{g_{ab}}{g_a} = 0, \tilde{\Delta} = 0.8$. Upper (lower) panels show density and phase for sublattice A(B).}
	\label{fig:l2_vortex_b}
\end{figure*}

The occupation of sublattice B can be enhanced by either opening a Haldane gap, which lowers the chemical potential on sublattice B, or by decreasing the intra-component repulsion on sublattice B, $\frac{g_b}{g_a} < 1$. In both cases, the increased kinetic energy is eventually compensated by the decreased potential energy, such that sublattice B supports a vortex, while a ring-soliton profiles is found on sublattice A. This phase is illustrated in figure \ref{fig:l2_vortex_b}. We call this configuration a B-vortex excitation.

\begin{figure*}
    \centering
    \includegraphics[width=0.9\textwidth]{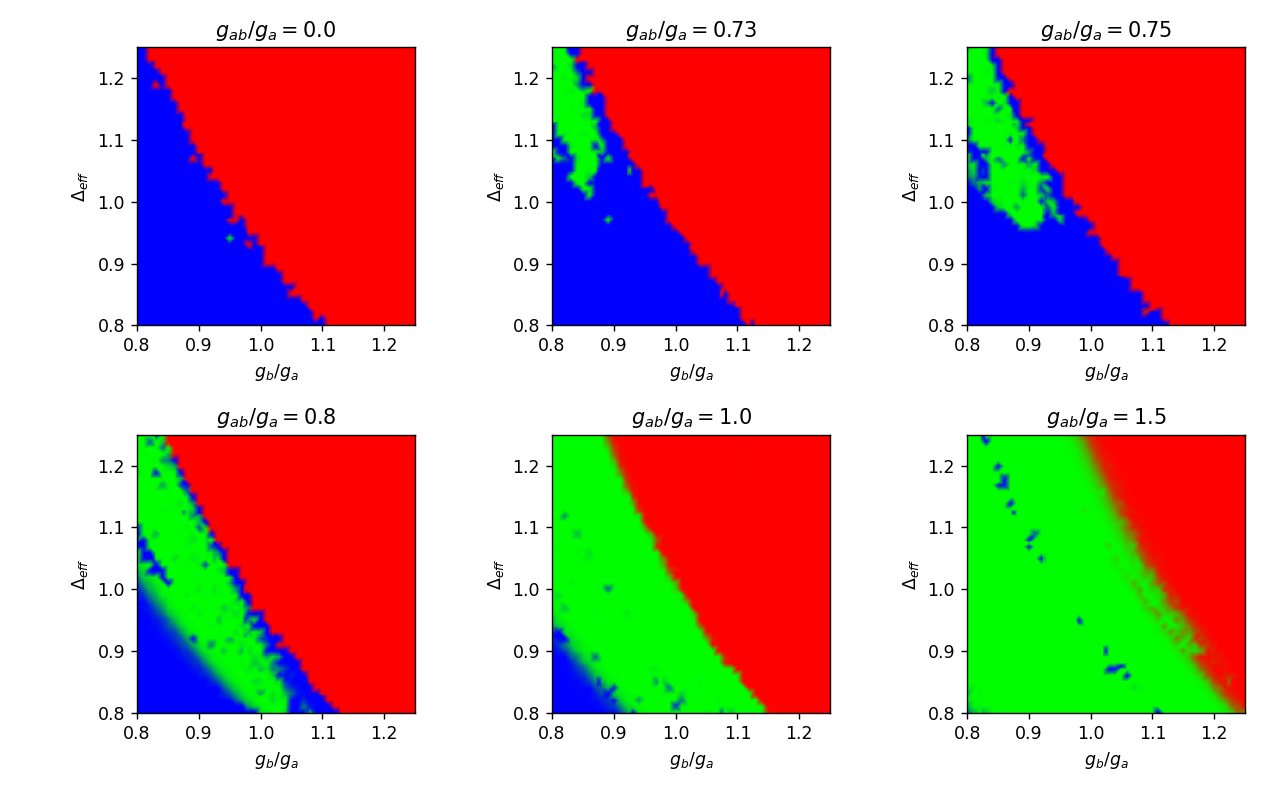}
    \caption{Phase diagram of vortex excitations for several values of $\frac{g_{ab}}{g_a}$. Red color indicates A-vortixes with winding number $l_a = 1$, blue color indicates B-vortices with winding number $l_b = 2$, and green color indicates a mixed phase with finite support on both components.}
    \label{fig:phase_diagram}
\end{figure*}

The resulting phase diagram as a function of the effective parameters $\frac{g_b}{g_a}$ and the effective Haldane gap $\tilde{\Delta}$ is shown in the upper left panel of Fig(\ref{fig:phase_diagram}, where the red color indicates an A-vortex, and the green color indicates a B-vortex.

\begin{figure*}
	\centering
	\includegraphics[width=0.9\textwidth]{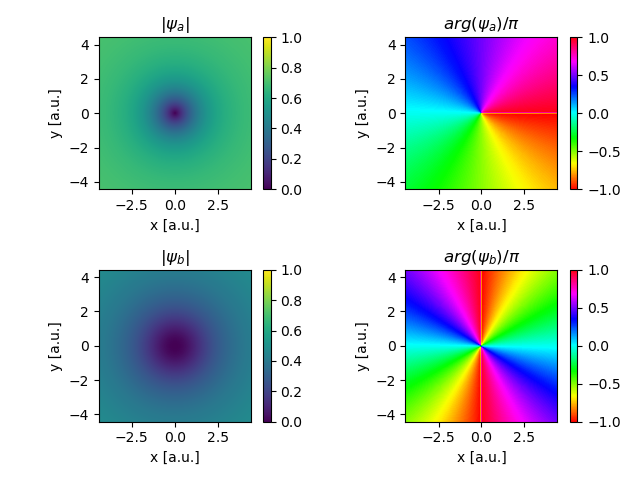}
	\caption{Normalized condensate density (left panels) and phase (right panels) of the multiple core vortex for $\ell_a=1,\ell_b=2,\frac{g_b}{g_a} = 1,\frac{g_{ab}}{g_a} = 1.1, \tilde{\Delta} = 1$. Upper (lower) panels show density and phase for sublattice A(B).}
	\label{fig:l2_vortex_both}
\end{figure*}

For sufficiently large inter-component interactions $g_{ab}$, we also observe a phase with co-existing vortex profiles on both components, shown in figure \ref{fig:l2_vortex_both}. Since the phase winding governs the vortex core behavior, $\rho(r) \sim r^\ell$, we find multiple core structures in this configuration. Introducing a Haldane gap and lowering the interaction strength $g_b$ facilitates the formation of a multiple core vortex. For dominant inter-component interactions, we find that the multiple core state is realized across a wide range of parameters.

We show these results in figure \ref{fig:phase_diagram}, where each panel shows a slice of the three-dimensional phase space spanned by {$\frac{g_b}{g_a}, \frac{g_{ab}}{g_a}, \tilde{\Delta}$} at fixed $\frac{g_{ab}}{g_a}$. Red color represents A-vortex states with winding number $\ell_a = 1$. Green color indicates the opposite B-vortex phase with $\ell_b=2$. Blue color indicates a multiple core state with simultaneous vortices on both components. The multiple core state first starts to appear at approximately $g_{ab} \approx 0.7 g_a$ for large Haldane gaps and $g_b < g_a$ (upper middle panel), and starts to replace the B-vortex state upon increasing $g_{ab}$, as indicated in the lower left panel. For dominant $g_{ab}$ the multiple core state is favored in large parts of the phase space, as shown in the lower right panel. These results represent the main findings of this paper. The visual artifacts in the phase diagram plots stem from the increased core size close to the phase boundaries, which leads to longer computation times and less reliable numerical detection of vortex states.

\subsection{$\ell_b = 1, \ell_a=0$ sector}

\begin{figure*}
    \centering
    \includegraphics[width=0.9\textwidth]{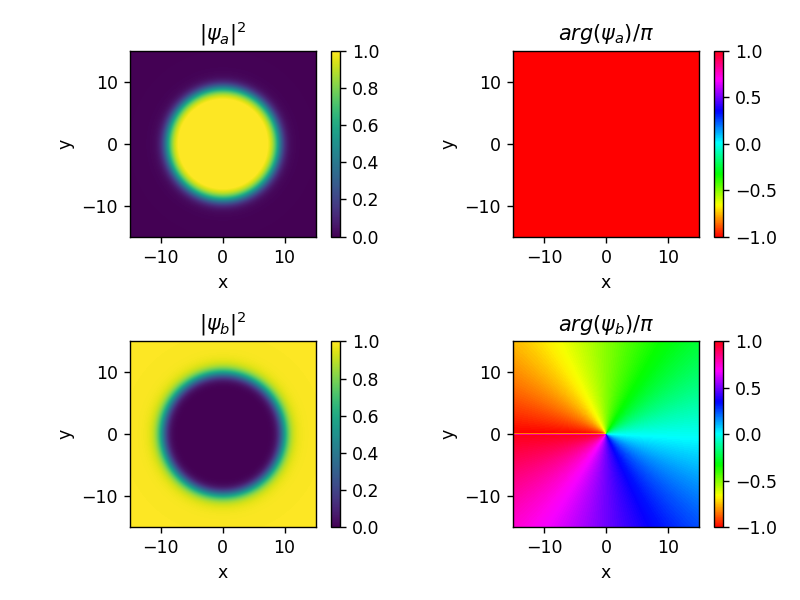}
    \caption{Normalized condensate density (left panels) and phase (right panels) of the B-vortex for $\ell_a=0,\ell_b=1$. Upper (lower) panels show density and phase for sublattice A(B).}
    \label{fig:l1_b_vortex}
\end{figure*}

Our main results generally hold for the topological sector $\ell_b = 1, \ell_a=0$ with slight modifications. We again identify three classes of solutions characterized by the far-field behaviors. Since sublattice A has no phase winding, the A-dominant phase is a homogeneous condensate on sublattice A, while sublattice B is unoccupied. The B-dominant phase typically shows a soliton-like switch between the sublattice, as shown in figure \ref{fig:l1_b_vortex}. In addition, we again observe co-existing vortices on both sublattices for large enough $\frac{g_{ab}}{g_a}$.

Generally, we observe that the core size is larger compared to the $\ell_b = 2,\ell_a=1$ sector, and therefore makes the numerical detection of vortex states more expensive and numerically unstable. We therefore omit the computation of the corresponding phase diagram.

\section{Conclusion} \label{sec:conclusion}
There have been numerous proposals to realize real and artificial materials hosting Bose excitations with topological nodal points, commonly known as bosonic Dirac and Weyl materials \cite{haddad1, haddad2, haddad3, haddad4, haddad5, haddad6, haddad7, sukhachov2020boseeinstein, Xu2018}. Combining this class of materials with non-equilibrium Bose-Einstein condensation gives access to the study of Dirac BEC, which are natural examples of multi-band condensates and feature a linear dispersion relation \cite{sukhachov2020boseeinstein}. We present vortex solutions in BEC of bosons with Dirac dispersion, Dirac BEC. Given the topological nature of the Dirac node, the phase winding of these condensates differ by one $|\ell_a-\ell_b| = 1$  \cite{haddad1}. Hence the kinetic and potential energies of two components is different, resulting in a different core structure. We  find that the far field behaviors for vortices develop different occupancies for pseudospin species (sublattices A and B). We present the detailed phase diagram for the the vortex states in Dirac BEC depending on the intra- and inter-component interaction strengths. We consider both the case of pure Dirac nodal points and the Dirac spectrum with a Haldane gap. The Haldane gap tends to favor the occupation of pseudo-spin B over A. 

A successful experimental observation of the Dirac BEC and vortices would represent an important step in the development of multi-band topological  condensates. The observation of vortices would depend on the physical system. For the specific realizations of the Dirac BEC we use the example of Dirac BEC in pumped Dirac magnons on Cr based trihalides. One would therefore expect a nontrivial features in magnon population and might use, e.g. the Brilloiun light scattering to observe Dirac BEC vortices. 

\section*{Acknowledgements}
We are grateful to G. Aeppli, J. Bailey, S. Banerjee and P. Suckachov for useful discussions. This work was supported by   the European Research Council under the European Union's Seventh Framework Program Synergy HERO-81451,  the Knut and Alice Wallenberg Foundation KAW 2018-0104, KAW 2019.0068 and by the University of Connecticut.

\section{Appendix: Numerical methods}
Here we outline the numerical methods used to obtain our results. First, we discuss the numerical shooting algorithm used to find vortex solutions of the radial GPE \eqref{gpe_radial}. Second, we discuss the bisection algorithm used to automatically scan the phase space for vortex solutions. Finally we outline the classification algorithm used to establish the phase diagram.

\subsection{Shooting algortihm}
Using a global phase rotation allows us to tune $\rho_a(r)$ to the real axis. For $l_b\geq2$, it then follows that $\rho_b$ is purely imaginary. Therefore, \eqref{gpe_radial} becomes a system of two coupled, non-linear differential equations, which may be expressed by the vector $\vec{u} = \begin{pmatrix} R_a & I_b \end{pmatrix}^T$:
\begin{widetext}
\begin{equation}
	\label{real_gpe}
	\vec{u}'(r) = \begin{pmatrix}
		\Big[\tilde{g}_b \big(u_2^2 - n_{b,\infty}\big) + \tilde{g}_{ab} u_1^2 \Big] u_2 - \frac{1-l_b}{r} u_1 \\
		-\Big[\tilde{g}_a \big(u_1^2 - n_{a,\infty}\big) + \tilde{g}_{ab} u_2^2 \Big] u_1 - \frac{l_b}{r} u_2
	\end{pmatrix}.
\end{equation}
\end{widetext}

Here, the coupling terms have been divided by the magnon velocity, $\tilde{g}_j = \frac{g_j}{v}$. Once the initial conditions are fixed, \eqref{real_gpe} can be integrated easily using forward integration. 

The initial conditions at $r=0$ are obtained from a Taylor expansion for $r\ll1$: Let $u_1(r) = \sum_{j\geq0} a_j r^j$ and $u_2(r) = \sum_{j\geq0} b_j r^j$. One then finds recursive relations for the coefficients,
\begin{widetext}
\begin{subequations}
	\begin{align}
		a_{j+1} (j+2-l) &= \tilde{g}_b\bigg( \sum_{l_1+l_2+l_3=j} b_{l_1} b_{l_2} b_{l_3} - n_{b,\infty} b_j \bigg) \nonumber \\ 
		&+ \tilde{g}_{ab} \sum_{l_1+l_2+l_3=j} a_{l_1} a_{l_2} b_{l_3}, \\
		b_{j+1} (j+1+l) &= - \tilde{g}_a\bigg( \sum_{l_1+l_2+l_3=j} a_{l_1} a_{l_2} a_{l_3} - n_{a,\infty} a_j \bigg) \nonumber \\
		&- \tilde{g}_{ab} \sum_{l_1+l_2+l_3=j} a_{l_1} b_{l_2} b_{l_3}.
	\end{align}
\end{subequations}
\end{widetext}

In addition, we find the conditions $a_0 = 0$ unless $l_b=1$ and $b_0 = 0$ unless $l_b=0$. It can be seen that the first non-zero coefficients are $a_{l_b-1}$ and $b_{l_b}$, hence $a_{l_b-1}$ is the only free parameter of the system. Indeed, all other coefficients follow from the recursion relations. 

A vortex solution can then be found by evaluating the behavior at large $r$: By requiring that the derivatives vanish in the far-field limit, we obtain either $u_{1}\rightarrow0$ or $u_1 \rightarrow \sqrt{n_{a,\infty} - \frac{g_{ab}}{g_a} u_2^2}$ resp. $u_2 \rightarrow \sqrt{n_{b,\infty} - \frac{g_{ab}}{g_b} u_1^2}$. It follows that a vortex solution is found by tuning the free parameter $a_{l-1}$ such that $u_{1,2}$ don't oscillate for large $r$. In practice, the integration range is bounded and there is a finite precision of the integration. Therefore, the onset of oscillatory behavior is the key indicator for a vortex solution: oscillations are pushed away further from the vortex core the closer the initial value is to $a_{l_b-1}^{vortex}$. Also, the forward integration must start at a small positive value $r_0 \ll 1$ in order to avoid the singularity at $r=0$. We observe a negligible dependence of the initial values $a_{l_b-1}^{vortex}$ on $r_0$, which we interpret as a signature of the numerical stability of our approach.

\subsection{Bisection algorithm}
\begin{figure*}
	\centering
	\includegraphics[width=0.9\textwidth]{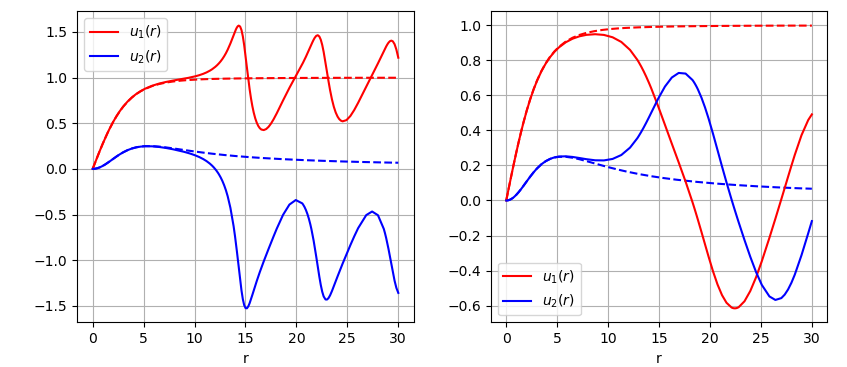}
	\caption{Convergence of the $\ell_a = 1, \ell_b=2$ vortex solution. Left panel: Overshooting behaviour for $a_1 > a_1^{vortex}$. Right panel: Undershooting behaviour for $a_1 < a_1^{vortex}$. The vortex solution obtained from the bisection algorithm is indicated by dashed lines.}
	\label{fig:l2_vortex_convergence}
\end{figure*}

To automate the numerical shooting, we use a bisection algorithm to search for the minimum number of oscillations of the integrated radial functions. Over- and undershooting behaviour can be distinguished by determining whether the integrated functions oscillate around the expected far-field limit or zero, respectively. An example of this behavior is shown in figure \ref{fig:l2_vortex_convergence}. It is clearly visible that overshooting (left panel) leads to oscillations around the far-field limits of each component, while undershooting (right panel) leads to a damped oscillation.

We remark that our algorithm has a number of limitations: First, the core size can become very large close to a phase boundary. We compensate for this by allowing an adaptive integration range, which is accompanied by a loss of numerical stability of the integration and higher computational cost. We therefore use a cutoff for the maximal integration range. Also, the core structure can contain extrema itself, which are hard to distinguish from oscillations in the recovered regime. Such behavior is typically also observed in the vicinity of a phase boundary. Second, the differential equation is usually very sensitive to the initial condition $a_{l-1}$ in the sense that small changes in $a_{l-1}$ drastically modify the far-field behavior. The bisection algorithm relies on this sensitivity to find the vortex solution. However, we observe that the sensitivity on $a_{l-1}$ is significantly reduced if both $\frac{g_b}{g_{ab}}$ and $\Delta_{eff}$ are large. Therefore the bisection algorithm is slowly convergent in this part of the phase space.

\subsection{Classification algorithm}
We employ a fuzzy clustering algorithm \cite{Dunn73} on the integrated radial profiles to establish the phase diagram. The fuzzy clustering algorithm is similar to k-means clustering, but allows for partial membership of a data point to each cluster, given by the membership matrix $W$. This is an advantage for the problem at hand, since the integrated wavefunctions can be ambiguous in the proximity of a phase boundary.

The algorithm minimizes the objective function

\begin{equation*}
    f(c) = \sum_{i=1}^{n} \sum_{j=1}^{n_c} W_{ij}^m \norm{x_i - c_j}^2,
\end{equation*}

for the data set {${x_1, \dots, x_n}$}, where $n_c$ is the number of clusters, $c$ is the array of centroids, and $m>1$ is the fuzzyness parameter. The centroids are determined by the "fuzzy" mean of the data set:

\begin{equation*}
    c_j = \frac{\sum_{i=1}^{n} W_{ij}^m x_i}{\sum_{i=1}^{n} W_{ij}^m}.
\end{equation*}

This is achieved in the following way: First, randomly assign values to the membership matrix and compute the centroids. For each iteration step, the membership matrix is updated according to
\begin{equation*}
    W_{ij} = \Bigg[ \sum_{k=1}^{n_c} \bigg(\frac{\norm{x_i - c_j}^2}{\norm{x_i - c_k}^2} \bigg)^{\frac{2}{m-1}}\Bigg]^{-1}.
\end{equation*}
Then update the centroids and repeat until convergence of the centroids is reached. The fuzzyness parameter $m$ governs how strict the membership of a data point to a single cluster will be; a larger value results in more fuzzyness.

For our case, we first identify the longest slowly-varying section of the integration range. This is necessary to exclude the vortex core for the analysis of the far-field behavior, and also the onset of oscillations at large $r$ present due to the numerical instability and finite accuracy of the initial condition $a_{l_b - 1}$. We then interpolate the slowly-varying section to ensure the same dimensionality for each data point. This allows us to compute the Euclidian distance norm between any two traces, which is the basis for the clustering algorithm.

\bibliography{references}

\begin{thebibliography}{10}

\bibitem{Wehling2014}
T.O. Wehling, A.M. Black-Schaffer, and A.V. Balatsky.
\newblock Dirac materials.
\newblock {\em Advances in Physics}, 63(1):1--76, 2014.

\bibitem{Lu2015}
Ling Lu, Zhiyu Wang, Dexin Ye, Lixin Ran, Liang Fu, John~D. Joannopoulos, and
  Marin Soljačić.
\newblock Experimental observation of weyl points.
\newblock {\em Science}, 349(6248):622--624, 2015.

\bibitem{Wang2016}
HaiXiao Wang, Lin Xu, HuanYang Chen, and Jian-Hua Jiang.
\newblock Three-dimensional photonic dirac points stabilized by point group
  symmetry.
\newblock {\em Phys. Rev. B}, 93:235155, Jun 2016.

\bibitem{Yang2019}
Yihao Yang, Zhen Gao, Haoran Xue, Li~Zhang, Mengjia He, Zhaoju Yang, Ranjan
  Singh, Yidong Chong, Baile Zhang, and Hongsheng Chen.
\newblock Realization of a three-dimensional photonic topological insulator.
\newblock {\em Nature}, 565(7741):622--626, Jan 2019.

\bibitem{kim_2013}
N~Y Kim, K~Kusudo, A~Löffler, S~Höfling, A~Forchel, and Y~Yamamoto.
\newblock Exciton{\textendash}polariton condensates near the dirac point in a
  triangular lattice.
\newblock 15(3):035032, mar 2013.

\bibitem{jacqmin_2014}
T.~Jacqmin, I.~Carusotto, I.~Sagnes, M.~Abbarchi, D.~D. Solnyshkov,
  G.~Malpuech, E.~Galopin, A.~Lema\^{\i}tre, J.~Bloch, and A.~Amo.
\newblock Direct observation of dirac cones and a flatband in a honeycomb
  lattice for polaritons.
\newblock {\em Phys. Rev. Lett.}, 112:116402, Mar 2014.

\bibitem{Samuelson71}
E.~J. Samuelsen, Richard Silberglitt, G.~Shirane, and J.~P. Remeika.
\newblock Spin waves in ferromagnetic cr${\mathrm{br}}_{3}$ studied by
  inelastic neutron scattering.
\newblock {\em Phys. Rev. B}, 3:157--166, Jan 1971.

\bibitem{Yelon71}
W.~B. Yelon and Richard Silberglitt.
\newblock Renormalization of large-wave-vector magnons in ferromagnetic
  cr${\mathrm{br}}_{3}$ studied by inelastic neutron scattering: Spin-wave
  correlation effects.
\newblock {\em Phys. Rev. B}, 4:2280--2286, Oct 1971.

\bibitem{Fransson2016}
J.~Fransson, A.~M. Black-Schaffer, and A.~V. Balatsky.
\newblock Magnon dirac materials.
\newblock {\em Phys. Rev. B}, 94:075401, Aug 2016.

\bibitem{pershoguba_2018}
Sergey~S. Pershoguba, Saikat Banerjee, J.~C. Lashley, Jihwey Park, Hans
  \AA{}gren, Gabriel Aeppli, and Alexander~V. Balatsky.
\newblock Dirac magnons in honeycomb ferromagnets.
\newblock {\em Phys. Rev. X}, 8:011010, Jan 2018.

\bibitem{Chen2018}
Lebing Chen, Jae-Ho Chung, Bin Gao, Tong Chen, Matthew~B. Stone, Alexander~I.
  Kolesnikov, Qingzhen Huang, and Pengcheng Dai.
\newblock Topological spin excitations in honeycomb ferromagnet
  ${\mathrm{cri}}_{3}$.
\newblock {\em Phys. Rev. X}, 8:041028, Nov 2018.

\bibitem{Xu2018}
Changsong Xu, Junsheng Feng, Hongjun Xiang, and Laurent Bellaiche.
\newblock Interplay between kitaev interaction and single ion anisotropy in
  ferromagnetic cri3 and crgete3 monolayers.
\newblock {\em npj Computational Materials}, 4(1):57, Nov 2018.

\bibitem{mcclarty_2018}
P.~A. McClarty, X.-Y. Dong, M.~Gohlke, J.~G. Rau, F.~Pollmann, R.~Moessner, and
  K.~Penc.
\newblock Topological magnons in kitaev magnets at high fields.
\newblock {\em Phys. Rev. B}, 98:060404, Aug 2018.

\bibitem{demokrit_2006}
So~Demokritov, V~Demidov, Oleksandr Dzyapko, G.~Melkov, Alexander Serga,
  Burkard Hillebrands, and Andrei Slavin.
\newblock Bose-einstein condensation of quasi-equilibrium magnons at room
  temperature under pumping.
\newblock {\em Nature}, 443:430--3, 10 2006.

\bibitem{demidov2007}
V.~E. Demidov, O.~Dzyapko, S.~O. Demokritov, G.~A. Melkov, and A.~N. Slavin.
\newblock Thermalization of a parametrically driven magnon gas leading to
  bose-einstein condensation.
\newblock {\em Phys. Rev. Lett.}, 99:037205, Jul 2007.

\bibitem{Demidov2008}
V.~E. Demidov, O.~Dzyapko, S.~O. Demokritov, G.~A. Melkov, and A.~N. Slavin.
\newblock Observation of spontaneous coherence in bose-einstein condensate of
  magnons.
\newblock {\em Phys. Rev. Lett.}, 100:047205, Jan 2008.

\bibitem{Serga2014}
Alexander~A. Serga, Vasil~S. Tiberkevich, Christian~W. Sandweg, Vitaliy~I.
  Vasyuchka, Dmytro~A. Bozhko, Andrii~V. Chumak, Timo Neumann, Bj{\"o}rn Obry,
  Gennadii~A. Melkov, Andrei~N. Slavin, and Burkard Hillebrands.
\newblock Bose--einstein condensation in an ultra-hot gas of pumped magnons.
\newblock {\em Nature Communications}, 5(1):3452, Mar 2014.

\bibitem{Nowik-Boltyk2012}
P.~Nowik-Boltyk, O.~Dzyapko, V.~E. Demidov, N.~G. Berloff, and S.~O.
  Demokritov.
\newblock Spatially non-uniform ground state and quantized vortices in a
  two-component bose-einstein condensate of magnons.
\newblock {\em Scientific Reports}, 2(1):482, Jun 2012.

\bibitem{Bozhko2016}
Dmytro~A. Bozhko, Alexander~A. Serga, Peter Clausen, Vitaliy~I. Vasyuchka,
  Frank Heussner, Gennadii~A. Melkov, Anna Pomyalov, Victor~S. L'vov, and
  Burkard Hillebrands.
\newblock Supercurrent in a room-temperature bose--einstein magnon condensate.
\newblock {\em Nature Physics}, 12(11):1057--1062, Nov 2016.

\bibitem{Bozhko2019}
Dmytro~A. Bozhko, Alexander J.~E. Kreil, Halyna~Yu. Musiienko-Shmarova,
  Alexander~A. Serga, Anna Pomyalov, Victor~S. L'vov, and Burkard Hillebrands.
\newblock Bogoliubov waves and distant transport of magnon condensate at room
  temperature.
\newblock {\em Nature Communications}, 10(1):2460, Jun 2019.

\bibitem{Borisenko2020}
I.~V. Borisenko, B.~Divinskiy, V.~E. Demidov, G.~Li, T.~Nattermann, V.~L.
  Pokrovsky, and S.~O. Demokritov.
\newblock Direct evidence of spatial stability of bose-einstein condensate of
  magnons.
\newblock {\em Nature Communications}, 11(1):1691, Apr 2020.

\bibitem{Eisenstein2004}
J.~P. Eisenstein and A.~H. MacDonald.
\newblock Bose--einstein condensation of excitons in bilayer electron systems.
\newblock {\em Nature}, 432(7018):691--694, Dec 2004.

\bibitem{Kasprzak2006}
J.~Kasprzak, M.~Richard, S.~Kundermann, A.~Baas, P.~Jeambrun, J.~M.~J. Keeling,
  F.~M. Marchetti, M.~H. Szyma{\'{n}}ska, R.~Andr{\'e}, J.~L. Staehli,
  V.~Savona, P.~B. Littlewood, B.~Deveaud, and Le~Si Dang.
\newblock Bose--einstein condensation of exciton polaritons.
\newblock {\em Nature}, 443(7110):409--414, Sep 2006.

\bibitem{balili2007}
R.~Balili, V.~Hartwell, D.~Snoke, L.~Pfeiffer, and K.~West.
\newblock Bose-einstein condensation of microcavity polaritons in a trap.
\newblock {\em Science}, 316(5827):1007--1010, 2007.

\bibitem{deng2010}
Hui Deng, Hartmut Haug, and Yoshihisa Yamamoto.
\newblock Exciton-polariton bose-einstein condensation.
\newblock {\em Rev. Mod. Phys.}, 82:1489--1537, May 2010.

\bibitem{bailey2022}
Joe Bailey, Pavlo Sukhachov, Korbinian Baumgaertl, Simone Finizio, Sebastian
  Wintz, Carsten Dubs, Joerg Raabe, Dirk Grundler, Alexander Balatsky, and
  Gabriel Aeppli.
\newblock Multi-band bose-einstein condensate at four-particle scattering
  resonance, 2022.

\bibitem{sukhachov2020boseeinstein}
P.~O. Sukhachov, S.~Banerjee, and A.~V. Balatsky.
\newblock Bose-einstein condensate of dirac magnons: Pumping and collective
  modes.
\newblock {\em Phys. Rev. Research}, 3:013002, Jan 2021.

\bibitem{Kourakis2005}
I.~Kourakis, P.~K. Shukla, M.~Marklund, and L.~Stenflo.
\newblock Modulational instability criteria for two-componentbose--einstein
  condensates.
\newblock {\em The European Physical Journal B - Condensed Matter and Complex
  Systems}, 46(3):381--384, Aug 2005.

\bibitem{haddad1}
L.~H. Haddad and L.~D. Carr.
\newblock Relativistic linear stability equations for the nonlinear dirac
  equation in bose-einstein condensates.
\newblock {\em {EPL} (Europhysics Letters)}, 94(5):56002, may 2011.

\bibitem{haddad2}
L.H. Haddad and L.D. Carr.
\newblock The nonlinear dirac equation in bose–einstein condensates:
  Foundation and symmetries.
\newblock {\em Physica D: Nonlinear Phenomena}, 238(15):1413 -- 1421, 2009.
\newblock Nonlinear Phenomena in Degenerate Quantum Gases.

\bibitem{haddad3}
L~H Haddad, C~M Weaver, and Lincoln~D Carr.
\newblock The nonlinear dirac equation in bose{\textendash}einstein
  condensates: I. relativistic solitons in armchair nanoribbon optical
  lattices.
\newblock {\em New Journal of Physics}, 17(6):063033, jun 2015.

\bibitem{haddad4}
L~H Haddad and Lincoln~D Carr.
\newblock The nonlinear dirac equation in bose{\textendash}einstein
  condensates: {II}. relativistic soliton stability analysis.
\newblock {\em New Journal of Physics}, 17(6):063034, jun 2015.

\bibitem{haddad5}
L~H Haddad and Lincoln~D Carr.
\newblock The nonlinear dirac equation in bose{\textendash}einstein
  condensates: superfluid fluctuations and emergent theories from relativistic
  linear stability equations.
\newblock {\em New Journal of Physics}, 17(9):093037, sep 2015.

\bibitem{haddad6}
L~H Haddad and Lincoln~D Carr.
\newblock The nonlinear dirac equation in bose{\textendash}einstein
  condensates: vortex solutions and spectra in a weak harmonic trap.
\newblock {\em New Journal of Physics}, 17(11):113011, oct 2015.

\bibitem{haddad7}
L.~H. Haddad, K.~M. O'Hara, and Lincoln~D. Carr.
\newblock Nonlinear dirac equation in bose-einstein condensates: Preparation
  and stability of relativistic vortices.
\newblock {\em Phys. Rev. A}, 91:043609, Apr 2015.

\bibitem{poddubny2018}
Alexander~N. Poddubny and Daria~A. Smirnova.
\newblock Ring dirac solitons in nonlinear topological systems.
\newblock {\em Phys. Rev. A}, 98:013827, Jul 2018.

\bibitem{cuevas2016}
Jes\'us Cuevas-Maraver, Panayotis~G. Kevrekidis, Avadh Saxena, Andrew Comech,
  and Ruomeng Lan.
\newblock Stability of solitary waves and vortices in a 2d nonlinear dirac
  model.
\newblock {\em Phys. Rev. Lett.}, 116:214101, May 2016.

\bibitem{BOUSSAID2016}
Nabile Boussaïd and Andrew Comech.
\newblock On spectral stability of the nonlinear dirac equation.
\newblock {\em Journal of Functional Analysis}, 271(6):1462--1524, 2016.

\bibitem{Dunn73}
J.~C. Dunn.
\newblock A fuzzy relative of the isodata process and its use in detecting
  compact well-separated clusters.
\newblock {\em Journal of Cybernetics}, 3(3):32--57, 1973.

\end{thebibliography}
\bibliographystyle{unsrt}
\end{document}